\documentclass[10pt, journal]{IEEEtran}
\usepackage[margin = 0.7in]{geometry}
\usepackage{amsmath}
\usepackage{algpseudocode}
\usepackage{amssymb}
\usepackage{amsfonts}
\usepackage{mathtools}
\usepackage{amscd}
\usepackage{mdframed}
\usepackage{mathrsfs}
\usepackage{graphicx}
\usepackage{bm}
\usepackage{url}
\usepackage{verbatim}
\usepackage[mathcal]{euscript}
\usepackage{booktabs}
\newtheorem{theorem}{Theorem}

\newtheorem{proposition}{Proposition}

\def\BibTeX{{\rm B\kern-.05em{\sc i\kern-.025em b}\kern-.08em
    T\kern-.1667em\lower.7ex\hbox{E}\kern-.125emX}}
    
\begin{document}
\title{On Opportunistic Selection of Common Randomness and LLR generation for Algebraic Group Secret-Key Generation}

\author{Rohit Joshi$^{\,*}$ and J. Harshan$^{\, * \, , \, \dagger}$ \\
$^{*}$Bharti School of Telecom Technology and Management, Indian Institute of Technology Delhi, India.\\
$^{\dagger}$Department of Electrical Engineering, Indian Institute of Technology Delhi, India.

\thanks{This work was supported by the Indigenous 5G Test Bed project from the Department of Telecommunications, Ministry of Communications, New Delhi, India.}
}
\maketitle
\begin{abstract}
It is well known that physical-layer key generation methods enable wireless devices to harvest symmetric keys by accessing the randomness offered by the wireless channels. Although two-user key generation is well understood, group secret-key (GSK) generation, wherein more than two nodes in a network generate secret-keys, still poses open problems. Recently, Manish Rao et al., have proposed the Algebraic Symmetrically Quantized GSK (A-SQGSK) protocol for a network of three nodes wherein the nodes share quantized versions of the channel realizations over algebraic rings, and then harvest a GSK.  Although A-SQGSK protocol guarantees confidentiality of common randomness to an eavesdropper, we observe that the key-rate of the protocol is poor since only one channel in the network is used to harvest GSK. Identifying this limitation, in this paper, we propose an opportunistic selection method wherein more than one wireless channel is used to harvest GSKs without compromising the confidentiality feature, thereby resulting in remarkable improvements in the key-rate.  Furthermore, we also propose a log-likelihood ratio (LLR) generation method for the common randomness observed at various nodes, so that the soft-values are applied to execute LDPC codes based reconciliation to reduce the bit mismatches among the nodes.
\end{abstract}


\section{Introduction} 

It is well known that the randomness offered by wireless channels can be exploited to harvest symmetric keys among the nodes in a wireless network \cite{R2}. In two-user physical-layer key generation, the nodes observe the temporal variation in their wireless channel by sharing probing signals with each other, followed by a consensus algorithm to synthesize a secret-key. However, in a more generic model of group secret-key (GSK) generation, more than two nodes intend to harvest a secret-key for securing communication in broadcast and relaying applications, e.g., in vehicular networks. Broadly, physical-layer GSK generation can be classified into two types: (i) Pairwise GSK generation, wherein a central authority (which is one of the nodes in the network) generates a secret-key by applying two-user key generation algorithm with one of its neighbours, and then shares it with the other nodes in the network in a confidential manner \cite{R3,R6}, and (ii) Group consensus based GSK generation, wherein the central authority assists multiple nodes in the network to witness a common source of randomness (CSR) so that all the nodes can synthesize a group secret-key using a group consensus algorithm \cite{R9,R1,R7,R8}. It is noted that the former scheme trusts the central authority in the process of key generation and key distribution whereas in the latter scheme all the nodes share the responsibility of synthesising the key through a group consensus algorithm on the observed CSR. One of the motivations for using the latter class of methods is that the group consensus algorithm, which is executed after sharing the CSR, allows the nodes to mitigate a number of insider attacks \cite{R5}, which could be executed by the central authority. In this paper, we are interested in the latter class of GSK generation protocols for the above mentioned reasons.

Recently, \cite{R9,R1} have proposed a group secret-key (GSK) protocol, referred to as the Algebraic Symmetrically Quantized GSK (A-SQGSK) for a three-user network, wherein one of the channels in the network, which is chosen as the CSR, is appropriately quantized over an algebraic ring, and then shared among the nodes while ensuring zero leakage of the CSR to an external eavesdropper. It is observed that the A-SQGSK protocol is the only protocol that preserves confidentiality \cite{R9} in the class of group consensus based GSK generation, and this feature is attributed to the use of algebraic rings. Although A-SQGSK protocol provides zero leakage, we point out that its key-rate is low because the protocol uses only one of the channels in the network as the CSR. Furthermore, in the context of two-user key generation, it is well known that reconciliation algorithms, e.g., using low-density parity check (LDPC) codes, can be used to arrive at zero mismatch rate among the keys at the nodes. Towards employing such reconciliation methods in A-SQGSK, we observe that aspects of generating log-likelihood ratios (LLRs) of the secret bits using the CSR samples have not been addressed hitherto. Identifying the above mentioned limitations, we make the following contributions in this work: 
\begin{enumerate}
\item For the framework of A-SQGSK protocol, we propose an opportunistic CSR selection strategy wherein the randomness offered by two channels are utilized to synthesize a GSK without compromising the confidentiality feature. We show that our approach provides higher key-rate than the protocol proposed in \cite{R9,R1}.
\item Furthermore, when both the channels are amenable to key generation, we emphasize that using both would lead to compromise in the confidentiality, and as a result, we present a likelihood based strategy to choose one of them so as to minimize the mismatch rate among the nodes. 
\item Finally, for the A-SQGSK protocol, we also propose an LLR generation scheme, using which all the nodes employ a reconciliation algorithm. We also use LDPC based reconciliation to validate the effectiveness of our LLR generation scheme. Although LLR generation on CSR samples for two-user key generation is well known, we highlight that the statistics of the underlying noise is different at different set of nodes in the A-SQGSK protocol \cite{R1}. Therefore, the proposed LLR generation scheme is a non-trivial contribution in GSK generation. 
\end{enumerate}

\section{System Model and Background}


\begin{figure}[htbp]
\centerline{\includegraphics[scale = 0.4]{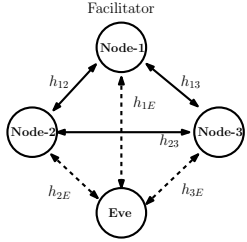}}
\vspace{-0.3cm}
\caption{\label{fig:one} A network of three nodes along with an eavesdropper }
\label{fig:model}
\end{figure}

A network comprising three nodes: node-1, node-2, and node-3 is considered, as shown in Fig. \ref{fig:model}. The channel between node-$j$ and node-$k$ is denoted by $h_{jk}$, where $ j \neq k$ and $h_{jk} \sim \mathcal{CN}(0,\,1)\,$.  Channel $h_{jk}$ is assumed: (i) to be flat-fading and remains quasi-static for a block of at least 4 channel uses, (ii) it exhibits pairwise reciprocity within the coherence-block  i.e., $h_{jk} = h_{kj}$ and (iii) $\{h_{jk}\}$ are statistically independent and identically distributed. Another assumption made is that all the nodes witness Additive White Gaussian Noise (AWGN) distributed  as $\mathcal{CN}(0,\,\sigma^2)\,$, so that the average signal-to-noise-ratio (SNR) is $\frac{1}{\sigma^2}$. To synthesize a GSK, a subset of $\{h_{12},h_{13}, h_{23}\}$, referred to as the CSR, must be learned by all the nodes. In \cite{R9}, the authors proposed the A-SQGSK scheme, wherein the channel $h_{12}$ is the chosen CSR among the nodes. While a noisy version of $h_{12}$ can be estimated at node-2 and node-1 by probing pilot symbols within a coherence-block, it is clear that node-3 needs to learn $h_{12}$ explicitly. To help this cause, \cite{R9} proposed a protocol to share quantized version of $h_{12}$ over an algebraic ring. First, we recall the A-SQGSK protocol, and then point out its limitations. 

\subsection{A-SQGSK Protocol}

\begin{figure*}
\begin{equation}
\label{proposition_2:eq_1}
I(C_1^{h_{12}},C_1^{h_{13}};\phi^{-1}(c_{sum}))= H(C_1^{h_{12}},C_1^{h_{13}}) -H(C_1^{h_{12}},C_1^{h_{13}}|\phi^{-1}(c_{sum})) 
\end{equation}
\begin{equation}
\vspace{-0.1cm}
\label{proposition_2:eq_2}
 H(C_1^{h_{12}},C_1^{h_{13}}) =  H(C_1^{h_{12}}) + H(C_1^{h_{13}}|C_1^{h_{12}})  = H(C_1^{h_{12}}) + H(C_1^{h_{13}}) 
\end{equation}
\begin{equation}
\vspace{-0.4cm}
\label{proposition_2:eq_3}
H(C_1^{h_{12}},C_1^{h_{13}}|\phi^{-1}(c_{sum}))  = 
\sum_{j=1}^{2^m} H(C_1^{h_{12}},C_1^{h_{13}}|\phi^{-1}(c_{sum})=c_j)P(\phi^{-1}(c_{sum})=c_j) 
\end{equation}
\begin{equation}
\vspace{-0.2cm}
\label{proposition_2:eq_4}
-\sum_{k=1}^{2^m}\sum_{l=1}^{2^m}P(C_1^{h_{12}}=b_k,C_1^{h_{13}}=a_l|\phi^{-1}(c_{sum})=c_j) log_2 (P(C_1^{h_{12}} , C_1^{h_{13}}=a_l|\phi^{-1}(c_{sum})=c_j))
\end{equation}
\begin{equation}
\vspace{-0.3cm}
\label{proposition_2:eq_5}
 P(C_1^{h_{12}}=b_k,C_1^{h_{12}}=\phi^{-1}( \phi(a_l) \ominus \phi(c_j) ))
=  
\begin{cases}
P(C_1^{h_{12}}=b_k) , & \text{if} \,\,\,b_k= \phi^{-1}( \phi(a_l) \ominus \phi(c_j) ) \\
0, & \text{otherwise}
\end{cases}
\end{equation}
\begin{equation}
\vspace{-0.2cm}
\label{proposition_2:eq_6}
H(C_1^{h_{12}},C_1^{h_{13}}|\phi^{-1}(c_{sum})=c_j)= -\sum_{k=1}^{2^m} P(C_1^{h_{12}}=b_k)log_2(P(C_1^{h_{12}}=b_k))
=H(C_1^{h_{12}}) 
\end{equation} 
\hrule
\end{figure*}

To execute the A-SQGSK protocol, the three nodes must be equipped with two complex constellations $\mathcal{A}'$ and $\bar{\mathcal{A}}$, as exemplified in Fig. \ref{fig:algebriac_transformation}. To formally define, $\mathcal{A}'$ is the algebraic ring $\mathbb{Z}_{2\frac{m}{2}}[i]$ where the set $\mathbb{Z}_{2^{\frac{m}{2}}}$, for some integer $m>1$, is given by $\{0,1,\ldots,2^{\frac{m}{2}}-1\}$. $\bar{\mathcal{A}}$ is a regular square quadrature amplitude modulation (QAM) constellation of size $2^{m}$, given by $\bar{\mathcal{A}} = \bar{\mathcal{A}}_{I} \bigoplus i\bar{\mathcal{A}}_{Q},$ such that $\bar{\mathcal{A}}_{I}$ = $\bar{\mathcal{A}}_{Q} = \{-2^{\frac{m}{2}} + 1, -2^{\frac{m}{2}} + 3, \ldots, 2^{\frac{m}{2}} - 3, 2^{\frac{m}{2}} - 1\}$, where $i = \sqrt{-1}$, and $m$ is even. The one-one transformation from $\bar{\mathcal{A}}$ to $\mathcal{A}'$, represented by $\phi: \bar{\mathcal{A}} \rightarrow \mathcal{A}'$, is $\phi(\alpha) = \frac{\alpha + 2^{\frac{m}{2}} - 1 + i(2^{\frac{m}{2}} - 1)}{2}
.$ The A-SQGSK protocol consists of four phases:

\begin{figure}
\begin{center}
\includegraphics[scale=0.46]{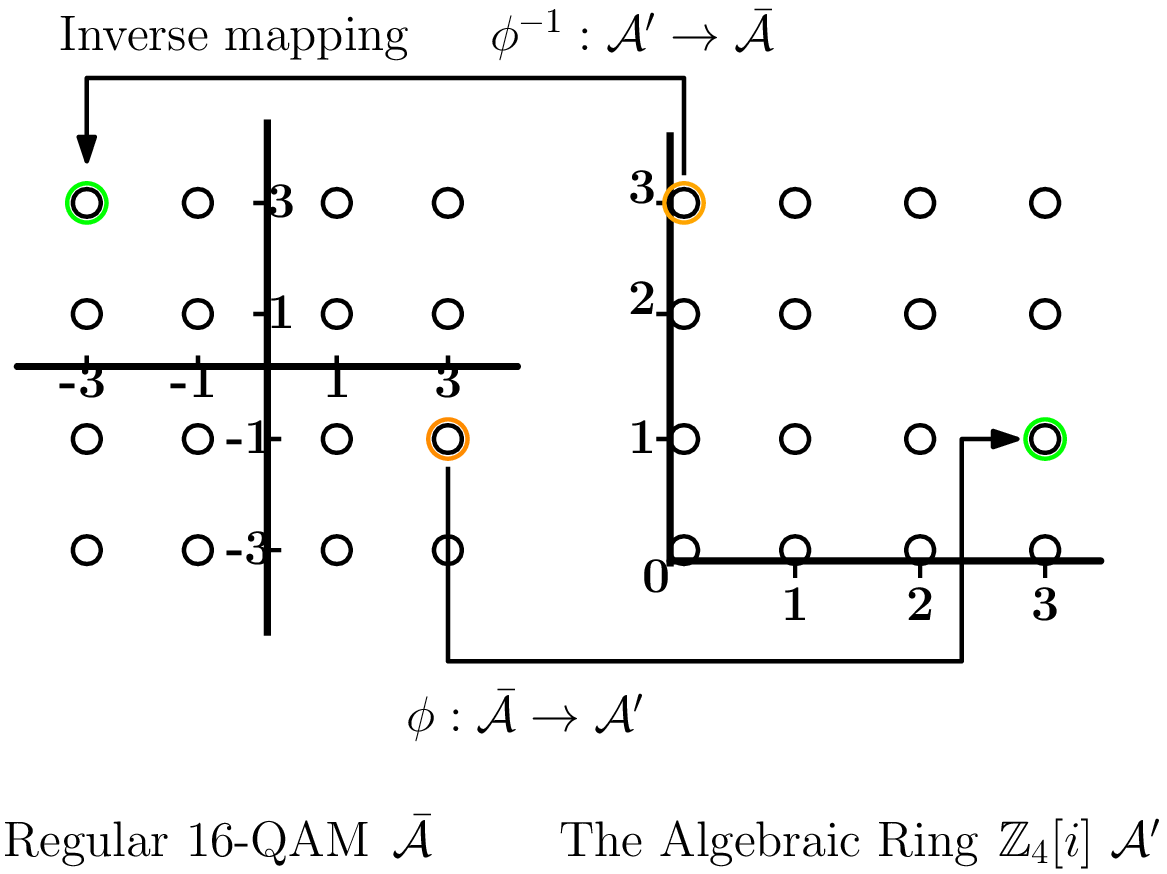}
\vspace{-0.2cm}
\caption{\label{fig:algebriac_transformation} Example for the two constellations to facilitate one-to-one transformation in the A-SQGSK protocol with $m = 4$.}
\end{center}
\end{figure}

\textbf{Phase-1}: node-1 broadcasts a pilot symbol $x=1$ using which node-2 and node-3 receive $y_2^{(1)}= h_{12}x+n_2^{(1)}$ and $y_3^{(1)}= h_{13}x + n_3^{(1)}$, where the noise $n_2^{(1)}$ and $n_3^{(1)}$ are distributed as $\mathcal{CN}(0,\,\sigma^2)\,$ respectively. In this notation, the superscript denotes the phase number in each coherence- block and the subscript denotes the node index. These two nodes then estimate the channel as $h_{12}+e_2^{(1)}$ and $h_{13}+e_3^{(1)}$, where the estimation errors are distributed as $e_2^{(1)} , e_3^{(1)}  \sim \mathcal{CN}(0,\,\gamma)\,$. Furthermore, these estimates are then quantized as 
\begin{equation*}
C^{h_{12}}_2 = \varphi(h_{12}+e_2^{(1)}) \in \bar{\mathcal{A}} \, \, \, \, \text{and }
\end{equation*}
\begin{equation*}
\hspace{-0.74cm}
C^{h_{13}}_{3} = \varphi(h_{13}+e_3^{(1)}) \in \bar{\mathcal{A}},
\end{equation*}
wherein in the superscript denotes the inherently observed channel and the subscript denotes the node index. Also, the quantization operator $\varphi (\beta)$ for $\beta \in \mathbb{C}$ is given as 
\begin{equation*}
\varphi (\beta) = \textit{arg~min}_{a\in \bar{\mathcal{A}}} |\beta - a|^2 \in \bar{\mathcal{A}},
\end{equation*}
where $\varphi (\cdot)$ works independently on in-phase and the quadrature components. 
\\\textbf{Phase-2}: Similarly,  node-2 broadcasts a pilot symbol $x=1$, which is used by node-1 and node-3 to estimate the channel as $h_{12}+e_1^{(2)}$ and $h_{23}+e_3^{(2)}$, respectively, with similar noise statistics as in \textbf{Phase-1}. Subsequently, the estimates are quantized as 
\begin{equation*}
C^{h_{12}}_{1} = \varphi(h_{12}+e_1^{(2)}) \in \bar{\mathcal{A}}  \, \, \, \, \text{and }
\end{equation*} 
\begin{equation*}
\hspace{-0.74cm}
C^{h_{23}}_{3} = \varphi(h_{23}+e_3^{(2)}) \in \bar{\mathcal{A}}.
\end{equation*}
\textbf{Phase-3}: Similarly, node-3 transmits a pilot symbol $x=1$, whereby node-1 and node-2 estimate the channel as $h_{13}+e_1^{(3)}$ and $h_{23}+e_2^{(3)}$, respectively. Subsequently, both the nodes obtain the quantized version of estimates as 
\begin{equation*}
C^{h_{13}}_{1} = \varphi(h_{13}+e_1^{(3)}) \in \bar{\mathcal{A}}  \, \, \, \, \text{and }
\end{equation*} 
\begin{equation*}
\hspace{-0.74cm}
C^{h_{23}}_{2} = \varphi(h_{23}+e_2^{(3)}) \in \bar{\mathcal{A}}.
\end{equation*}
\textbf{Phase-4}: Assuming that the CSR is derived using $h_{12}$, node-3 does not have the access to it. To bridge the gap, node-1 computes the sum $c_{sum}=\phi(C^{h_{12}}_{1}) \oplus\phi(C^{h_{13}}_{1}) \in \mathcal{A}'$ over the ring $\mathbb{Z}_{2\frac{m}{2}}[i]$ and then it broadcasts $\frac{1}{\sqrt{E_{avg}}}\phi^{-1}(c_{sum})$ to node-2 and node-3, where $E_{avg}$ is the average energy of the QAM constellation. 

With the knowledge of $h_{13} + e^{(1)}_3$, node-3 obtains the \textit{maximum aposteriori probability} (MAP) estimate of $\phi^{-1}(c_{sum})$, denoted by $\hat{\theta}_3 \in \bar{\mathcal{A}}$. Using the above estimate, node-3 obtains the CSR as 
\begin{equation*}
C_3^{h_{12}}=  \phi^{-1}\bigg(\phi(\hat{\theta}_3)\ominus \phi(\varphi(h_{13} + e^{(1)}_3))\bigg) \in \bar{\mathcal{A}},
\end{equation*}
 where $\ominus$ is the subtraction over the ring $\mathbb{Z}_{2\frac{m}{2}}[i]$. Similarly, node-2 will recover $C_2^{h_{13}}$, which is the decoded version of the quantized channel $h_{13}$. By the end of the A-SQGSK protocol, node-$j$ has $\{C^{h_{12}}_{j}, C^{h_{13}}_{j} \}$ for $j = 1, 2, 3$. node-$j$ unfolds the real and the imaginary components of $C^{h_{12}}_{j}$, $C^{h_{13}}_{j}$, and then uses the samples for key extraction. Henceforth, throughout this paper, we refer to a CSR from the unfolded set of $h_{12}$ as $R^{h_{12}}_{j}$, and similarly, we refer to a CSR sample from the unfolded set of $h_{13}$ as $R^{h_{13}}_{j}$.

\subsection{Consensus Phase}
In order to extract the secret-key, \cite{R9} proposed to run the A-SQGSK protocol for a number of coherence-blocks, and then used $\{R^{h_{12}}_{j}\}$ as the CSR at node-$j$. Subsequently, a two-level consensus algorithm \cite{R2}, with guard bands $q_{+} \geq 0$, and $q_{-} \leq 0$, was employed to synthesize secret bits by satisfying the rule $\mathcal{Q}(\alpha) = 1$ if $\alpha > q_{+}$, and $\mathcal{Q}(\alpha) = 0$ if $\alpha < q_{-}$, for any real sample $\alpha$. A sample $\alpha$ is said to be out of consensus if $q_{-} \leq \alpha \leq q_{+}$. The guards bands were appropriately chosen to upper bound the mismatch rate (referred to as initial error rate), which is the fraction of bits that do not agree between any two nodes. To achieve consensus among the three nodes, \cite{R9} proposed all the three nodes to parse through their quantized samples of $\{R^{h_{12}}_{1}, R^{h_{12}}_{2}, R^{h_{12}}_{3}\}$, and then create a list of all the CSR samples that are lying outside the guard bands. Subsequently, all the nodes mutually agree on common indices and then generate the secret-key using the samples on the common indices. 

\begin{figure*}
\begin{small}
\begin{equation}
\label{eq:quantize_pdf}
\mho(\mathcal{N}(\mu, \gamma), \bar{\mathcal{A}}_{I}) = \left\lbrace \int_{-\infty}^{\bar{\mathcal{A}}_{I}(1) + \frac{d_{min}}{2}} P_{\Theta}(\theta) d\theta, \int_{\bar{\mathcal{A}}_{I}(2) - \frac{d_{min}}{2}}^{\bar{\mathcal{A}}_{I}(2) + \frac{d_{min}}{2}} P_{\Theta}(\theta) d\theta, \ldots, \int_{\bar{\mathcal{A}}_{I}(2^{\frac{m}{2}}) - \frac{d_{min}}{2}}^{\infty} P_{\Theta}(\theta) d\theta\right\rbrace
\end{equation}
\begin{equation}
\label{equation_LLR_selection_1}
\mbox{Prob}\left(R^{h_{13}}_{2} \in \mathcal{S} | R^{h_{13}}_{1} \in \bar{\mathcal{S}} \right) = \sum_{x_{u} \in \mathcal{S}} \varrho^{x_{u}}\left( \varnothing^{\phi(R^{h_{12}}_{1})} \left(\mho\left(\mathcal{N}\left(\phi^{-1} \bigg(\phi(R^{h_{12}}_{1}) \oplus \phi(R^{h_{13}}_{1})\bigg),\frac{E_{avg} \sigma^{2}}{|h_{12} + e^{(2)}_1|^2}\right), \bar{\mathcal{A}}_{I}\right) \right)  \right)
\end{equation}
\begin{equation}
\label{equation_LLR_selection_2}
\mbox{Prob}\left(R^{h_{12}}_{3} \in \mathcal{S} | R^{h_{12}}_{1} \in \bar{\mathcal{S}} \right) = \sum_{x_{u} \in \mathcal{S}} \varrho^{x_{u}}\left( \varnothing^{\phi(R^{h_{13}}_{1})} \left(\mho\left(\mathcal{N}\left(\phi^{-1} \bigg(\phi(R^{h_{12}}_{1}) \oplus \phi(R^{h_{13}}_{1})\bigg),\frac{E_{avg} \sigma^{2}}{|h_{13} + e^{(3)}_1|^2}\right), \bar{\mathcal{A}}_{I}\right) \right)  \right)
\end{equation}
\end{small}
\hrule
\end{figure*}

\section{Opportunistic CSR Selection}

\begin{figure}[htbp]
\centerline{\includegraphics[scale = 0.37]{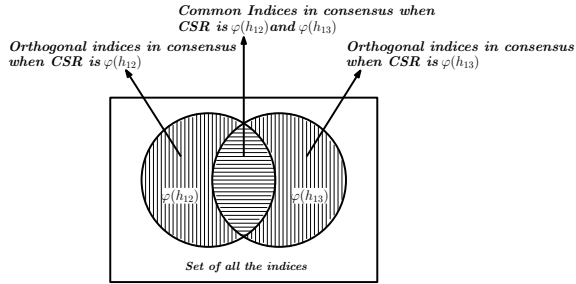}}
\vspace{-0.3cm}
\caption{\label{fig:one} Venn diagram depicting two possibilities of channels in consensus - (i) indices belongs to only one type of CSR, either $\varphi(h_{12})$ or $\varphi(h_{13})$, and (ii) indices belong to both the CSR (shown in the intersection).}
\label{fig:venn_diagram}
\end{figure}

One of the limitations of \cite{R9} is that only the CSR $\{R^{h_{12}}_{1}, R^{h_{12}}_{2}, R^{h_{12}}_{3}\}$ was used to extract secret-keys. However, as depicted on the right-side of Fig. \ref{fig:venn_diagram}, it is clear that the three nodes can opportunistically make use of both $\{R^{h_{12}}_{1}, R^{h_{12}}_{2}, R^{h_{12}}_{3}\}$ and $\{R^{h_{13}}_{1}, R^{h_{13}}_{2}, R^{h_{13}}_{3}\}$ based on the coherence-block under consideration. In particular, the following possibilities arise on a given coherence-block: (i) Only one of the two sets of CSR samples is out of the guard band at all the three nodes, thereby contributing to the key. (ii) Both the CSR sets lie in the guard band of at least one of the nodes, thereby not contributing to the key for this coherence block. (iii) Both the CSR sets lie outside the guard band at all the nodes, and therefore, either of them is a good choice of CSR. In the first case, the nodes can use the CSR which is in consensus on a given coherence-block. As a result, there will be improvement in the key-rate in comparison with the A-SQGSK protocol \cite{R9}. We prove that using either of the subset as the CSR preserves confidentiality.
\begin{proposition}
\label{prop1}
For a $2^m$-QAM constellation, when the two CSR, $C_1^{h_{12}}$ and $C_1^{h_{13}}$, are identically distributed, we have $I(C_1^{h_{12}};\phi^{-1}(c_{sum}))=0$ and $I(C_1^{h_{13}};\phi^{-1}(c_{sum}))=0$, where $\phi^{-1}(c_{sum})$ is the symbol transmitted by node-1.
\end{proposition}
\begin{IEEEproof}
Following the similar lines of the proof in \cite[Theorem 1]{R1}, it can be proved that $I(C_1^{h_{12}};\phi^{-1}(c_{sum}))=0$ and likewise, $I(C_1^{h_{13}};\phi^{-1}(c_{sum}))=0$. 
\end{IEEEproof}

Using Proposition \ref{prop1}, it follows that choosing either of the two CSR sets for a coherence block will not compromise the confidentiality feature of the CSR. As seen in the third case, on a given coherence-block, when both the CSR samples are in consensus, we cannot use both to extract the keys as it does not ensure confidentiality of the CSR samples as we prove next. 
\begin{proposition}
\label{prop2}
For a $2^m$-QAM constellation, when the two complex CSR, $C_1^{h_{12}}$ and $C_1^{h_{13}}$, are identically distributed, we have non-zero value of $I(C_1^{h_{12}},C_1^{h_{13}};\phi^{-1}(c_{sum}))$, where $\phi^{-1}(c_{sum})$ is the symbol transmitted by node-1.
\end{proposition} 
\begin{IEEEproof}
The expression for $I(C_1^{h_{12}},C_1^{h_{13}};\phi^{-1}(c_{sum}))$ is expanded in \eqref{proposition_2:eq_1}, where $H(C_1^{h_{12}},C_1^{h_{13}})$ is given in \eqref{proposition_2:eq_2} such that the second equality holds as $C_1^{h_{12}}$ and $C_1^{h_{13}}$ are statistically independent. Furthermore, $H(C_1^{h_{12}},C_1^{h_{13}}|\phi^{-1}(c_{sum}))$ is given in \eqref{proposition_2:eq_3}. \eqref{proposition_2:eq_4} to \eqref{proposition_2:eq_6} show that the conditional entropy, $H(C_1^{h_{12}},C_1^{h_{13}}|\phi^{-1}(c_{sum}))$, is equal to $H(C_1^{h_{12}})$ as follows. The first terms after summation in \eqref{proposition_2:eq_3}, $H(C_1^{h_{12}},C_1^{h_{13}}|\phi^{-1}(c_{sum})=c_j)$, is given in \eqref{proposition_2:eq_4}. The probability, $P(C_1^{h_{12}}=b_k,C_1^{h_{13}}=a_l|\phi^{-1}(c_{sum})=c_j)$ in \eqref{proposition_2:eq_4} is given as  $P(C_1^{h_{12}}=b_k,C_1^{h_{12}}=\phi^{-1}( \phi(a_l) \ominus \phi(c_j) ))$ in \eqref{proposition_2:eq_5}. Substituting \eqref{proposition_2:eq_5} in \eqref{proposition_2:eq_4} gives $H(C_1^{h_{12}},C_1^{h_{13}}|\phi^{-1}(c_{sum})=c_j)$ in \eqref{proposition_2:eq_6} which is equal to $H(C_1^{h_{12}})$  and then substituting \eqref{proposition_2:eq_6} in \eqref{proposition_2:eq_3} and \eqref{proposition_2:eq_3} in \eqref{proposition_2:eq_1} gives  $I(C_1^{h_{12}},C_1^{h_{13}};\phi^{-1}(C_{sum}))=H(C_1^{h_{13}})$.
\end{IEEEproof}

Assuming Eve can perfectly retrieve $\phi^{-1}(c_{sum})$, the leakage at Eve is non-zero using  Proposition \ref{prop2}.
Therefore, when $\{R^{h_{12}}_{1}, R^{h_{12}}_{2}, R^{h_{12}}_{3}\}$ and $\{R^{h_{13}}_{1}, R^{h_{13}}_{2}, R^{h_{13}}_{3}\}$ are in consensus on a given coherence-block, we present a method for selecting one of them such that the mismatch rate among the keys is minimized.

\subsection{Likelihood Based CSR selection Strategy}

\begin{figure}[htbp]
\centerline{\includegraphics[scale = 0.4]{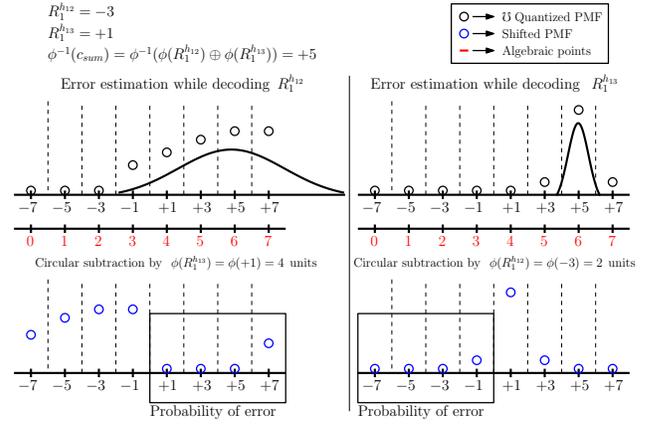}}
\vspace{-0.2cm}
\caption{\label{fig:two} Figure depicts an example for likelihood based CSR selection at node-1. On the left: an illustration of the computation of probability of error when the CSR is $R_{1}^{h_{12}}$. On the right: an illustration of the computation of probability of error when the CSR is $R_{1}^{h_{13}}$.
}
\label{fig:no_error}
\end{figure}

We present an optimal CSR selection strategy, wherein the facilitator first builds the likelihood functions on the CSR observed at node-2 and node-3, and then chooses the one that provides smaller probability of error. First, we define notations needed to explain the technique. For a Gaussian probability density function (PDF) $P_{\Theta}(\theta)$, denoted by $\mathcal{N}(\mu, \gamma)$ with mean $\mu$ and variance $\gamma$, the notation $\mho(\mathcal{N}(\mu, \gamma), \bar{\mathcal{A}}_{I})$ represents the probability mass function (PMF) induced on the discrete constellation $\bar{\mathcal{A}}_{I}$ when quantizing $P_{\Theta}(\theta)$ onto the points in $\bar{\mathcal{A}}_{I}$. In other words, $\mho(\mathcal{N}(\mu, \gamma), \bar{\mathcal{A}}_{I})$ is given in \eqref{eq:quantize_pdf}, where $d_{min}$ represents the minimum Euclidean distance of the constellation $\bar{\mathcal{A}_{I}}$, and $\bar{\mathcal{A}_{I}}(t)$ denotes the $t$-th component for $1 \leq t \leq 2^{\frac{m}{2}}$. We use $\varnothing^{s}(\mathbf{g})$ to denote circular shift of the elements of the vector $\mathbf{g}$ to the left by $s$ units. For a given PMF $\mathcal{H}$ on $\bar{\mathcal{A}}_{I}$, the notation $\varrho^p(\mathcal{H})$ denotes the probability of the sample point $p \in \bar{\mathcal{A}}_{I}$. Let $\hat{\bar{\mathcal{A}}}^-_{I}$ and $\hat{\bar{\mathcal{A}}}^+_{I}$ denote the set of PAM points in $\bar{\mathcal{A}}_{I}$ that are out of the guard band on the negative and positive sides, respectively.

\begin{theorem} On a coherence-block when both $\{R^{h_{12}}_{1}, R^{h_{12}}_{2}, R^{h_{12}}_{3}\}$ and $\{R^{h_{13}}_{1}, R^{h_{13}}_{2}, R^{h_{13}}_{3}\}$ are in consensus, the CSR of interest must be $\{R^{h_{12}}_{1}, R^{h_{12}}_{2}, R^{h_{12}}_{3}\}$ if $\mbox{Prob}\left(R^{h_{13}}_{2} \in \mathcal{S} | R^{h_{13}}_{1} \in \bar{\mathcal{S}} \right) \geq \mbox{Prob}\left(R^{h_{12}}_{3} \in \mathcal{S} | R^{h_{12}}_{1} \in \bar{\mathcal{S}} \right),$ or $\{R^{h_{13}}_{1}, R^{h_{13}}_{2}, R^{h_{13}}_{3}\}$ otherwise, where $\mbox{Prob}\left(R^{h_{13}}_{2} \in \mathcal{S} | R^{h_{13}}_{1} \in \bar{\mathcal{S}} \right)$ and $\mbox{Prob}\left(R^{h_{12}}_{3} \in \mathcal{S} | R^{h_{12}}_{1} \in \bar{\mathcal{S}} \right)$ are given in \eqref{equation_LLR_selection_1} and \eqref{equation_LLR_selection_2}, respectively. In this context, we have $\mathcal{S} = \hat{\bar{\mathcal{A}}}^-_{I}$, when $\bar{\mathcal{S}} = \hat{\bar{\mathcal{A}}}^+_{I}$. Similarly, we have $\mathcal{S} = \hat{\bar{\mathcal{A}}}^+_{I}$, when $\bar{\mathcal{S}} = \hat{\bar{\mathcal{A}}}^-_{I}$.
\end{theorem} 
\begin{IEEEproof}
Let us consider a coherence-block for which both $R^{h_{12}}_{j}$ and $R^{h_{13}}_{j}$ are in consensus for each $j \in \{1, 2, 3\}$. The corresponding quantized versions of the complex channels are $C^{h_{12}}_{j}$ and $C^{h_{13}}_{j}$. At sufficiently large SNR values, and an appropriate value of $m$, we have $C^{h_{12}}_{1} = C^{h_{12}}_{2}$ and $C^{h_{13}}_{1} = C^{h_{13}}_{3}$ with high probability. Using $R^{h_{12}}_{1}$ and $R^{h_{13}}_{1}$, node-1 generates the point that was broadcast to node-2 and node-3 as 
$\mu = \phi^{-1} \bigg(\phi(R^{h_{12}}_{1}) \oplus \phi(R^{h_{13}}_{1})\bigg)$. Using $h_{12} + e^{(2)}_1$ as an estimate of the channel seen between node-$1$ and node-$2$, node-$1$ builds an aposteriori PDF at node-2, and in this case it is Gaussian distributed given by $\mathcal{N}\left(\mu, \frac{E_{avg} \sigma^2}{|h_{12} + e^{(2)}_1|^2}\right)$. Furthermore, since node-2 decodes on the PAM constellation using MAP decoder, the corresponding aposteriori PMF on the PAM points is given by $\mathcal{H}_{R^{h_{13}}_{2}} = \mho\left(\mathcal{N}\left(\mu, \frac{E_{avg} \sigma^2}{|h_{12} + e^{(2)}_1|^2}\right), \bar{\mathcal{A}}\right)$, wherein the PMF $\mathcal{H}_{R^{h_{13}}_{2}}$ is listed on PAM points when enumerated in the increasing order. Finally, since node-2 obtains the CSR $R^{h_{13}}_{2}$ by performing a modulo subtraction on the algebraic ring, node-$1$ incorporates the corresponding changes in the PMF as $\bar{\mathcal{H}}_{R^{h_{13}}_{2}} = \varnothing^{\phi(R^{h_{12}}_{1})}(\mathcal{H}_{R^{h_{13}}_{2}})$. Thus, node-$1$ generates an aposteriori PMF on the CSR $R^{h_{13}}_{2}$ seen at node-$2$. Once the PMFs are generated, then the probability of error at node-2 is computed by summing over the mass points in the complementary region of the PAM constellation with respect to the CSR $R^{h_{13}}_{1}$. By mimicking similar operations at node-3, node-$1$ also generates $\bar{\mathcal{H}}_{R^{h_{12}}_{3}}$, which is an aposteriori PMF on the CSR $R^{h_{12}}_{3}$ seen at node-$3$, and then computes the probability of error at node-3. Finally, the CSR that provides lower probability of error is chosen for key generation. 
\end{IEEEproof}

Fig. \ref{fig:two} depicts an example for the likelihood selection strategy at node-1 when both the CSR are in consensus. We highlight that node-1 is able to generate the aposteriori PMFs seen at node-2 and node-3 by using the channel realizations available in the first four phases of the A-SQGSK protocol. As a result, no additional communication-overheads are involved. Furthermore, this method is optimal at moderate values of $m$ and mid-to-high SNR values since the quantized values of the channels used at node-1 would be the same used at node-2 and node-3 with high probability. 

\subsection{Consensus Algorithm for Opportunistic Selection}
\label{consensus_algorithm}
After executing the A-SQGSK protocol over $L$ coherence-blocks, the three nodes have a sequence of samples. For $r \in \{2, 3\}$, a CSR sample is said to come from $\varphi(h_{1r})$ if it is obtained from either the real or the imaginary part of the quantized version of $h_{1r}$ on any coherence-block. 
To achieve consensus, the three nodes use a generalized version of the consensus algorithm in \cite{R2} as follows: node-2 obtains two sets of indices, which comprises index values of the CSR samples of $\varphi(h_{12})$ and $\varphi(h_{13})$ lying outside the guard band, and then shares it to node-1. Upon receiving the indices, node-1 computes the corresponding sets of indices in a similar fashion for the two sets of CSR samples, and then broadcasts the set of indices that are in consensus with node-2. node-3 computes the corresponding sets of indices lying outside the guard band, and then broadcasts the two sets of indices that are in consensus with both node-1 and node-2.
Let $(\mathcal{R}_{\varphi(h_{12})} ,\mathcal{R}_{\varphi(h_{13})})$ denote the two sets of indices in consensus among the three nodes.  Then, all the three nodes generate the set $\mathcal{V}= \mathcal{R}_{\varphi(h_{12})} \cap \mathcal{R}_{\varphi(h_{13})}$, where $\mathcal{V}$ denotes the indices where both $\{R^{h_{12}}_{1}, R^{h_{12}}_{2}, R^{h_{12}}_{3}\}$ and $\{R^{h_{13}}_{1}, R^{h_{13}}_{2}, R^{h_{13}}_{3}\}$ are in consensus. With the likelihood based CSR selection strategy, for the indices in $\mathcal{V}$, node-1 calculates the probability of errors at node-2 and node-3 by locally generating the distributions at their side. Then it broadcasts the index of the chosen CSR to both node-2 and node-3. Finally, all the nodes use the CSR samples of $\varphi(h_{12})$ and $\varphi(h_{13})$ to extract a secret-key.

\begin{figure*}
\begin{small}
\begin{equation}
\vspace{-0.05cm}
\label{equation_theorem:LLR_decoding_1}
\mbox{Prob}_{b}(R^{h_{13}}_{2})= \sum_{x_{u} \in \mathcal{S}} \mbox{Prob}(x_{u})\left(\varrho^{R^{h_{13}}_{2}}\left( \varnothing^{\phi(R^{h_{12}}_{2})} \left(\mho\left(\mathcal{N}\left(\phi^{-1}(\phi(x_{u}) \oplus \phi(R^{h_{12}}_{2})\right),\frac{E_{avg} \sigma^{2}}{|h_{12} + e^{(1)}_2|^2}), \bar{\mathcal{A}}_{I}\right) \right)  \right) \right)
\end{equation}
\end{small}
\hrule
\end{figure*}

\section{LLR Based Reconciliation with LDPC codes}

We present an optimal LLR generation scheme for the CSR generated by the A-SQGSK protocol.  
Since node-1 observes the CSR samples from $\varphi(h_{12})$ and $\varphi(h_{13})$ through probing signals, we use the key generated at node-1 as the reference key, and then apply the LDPC reconciliation algorithm at node-2 and node-3. As the statistics of the underlying noise are different, the LLR generation scheme depends on whether the reconciliation is implemented at the (i) reciprocal node, which is the node that inherently observes the CSR through channel reciprocity, e.g., node-2 when the CSR is $\{R^{h_{12}}_{1}, R^{h_{12}}_{2}, R^{h_{12}}_{3}\}$, or the (ii) decoding node, which is the node that learns the unseen CSR through the process of decoding and subtraction over the ring, e.g., node-2 when the CSR is $\{R^{h_{13}}_{1}, R^{h_{13}}_{2}, R^{h_{13}}_{3}\}$. We discuss the LLR generation scheme at both these types of nodes.

\subsection{LLR Generation at the Reciprocal Node}
Let $P(X,Y)$ denote the joint PMF between the CSR at node-1, denoted by $X$, and the CSR at the reciprocal node, denoted by $Y$. For all the $2^{m}$ points in $\bar{\mathcal{A}}_I$, the probability that the CSR sample at node-1 is quantized to bit 1 and bit 0 is given by $\mbox{Prob}_1(p)= \sum_{ X \in \bar{\mathcal{A}}_I^+} P(X,Y=p) $
and
$\mbox{Prob}_0(p)= \sum_{ X \in \bar{\mathcal{A}}_I^-} P(X ,Y = p),$ respectively, 
where $\bar{\mathcal{A}}_I^+$ and $\bar{\mathcal{A}}_I^-$ are the positive and negative points of $\bar{\mathcal{A}}_I$, respectively, and $p$ is the CSR sample observed at the reciprocal node. Finally, the LLR is computed as $\mbox{log}\left(\frac{\mbox{Prob}_0(p)}{\mbox{Prob}_1(p)}\right)$.


\subsection{LLR Generation at the Decoding Node}

\begin{figure}[htbp]
\centerline{\includegraphics[scale = 0.5]{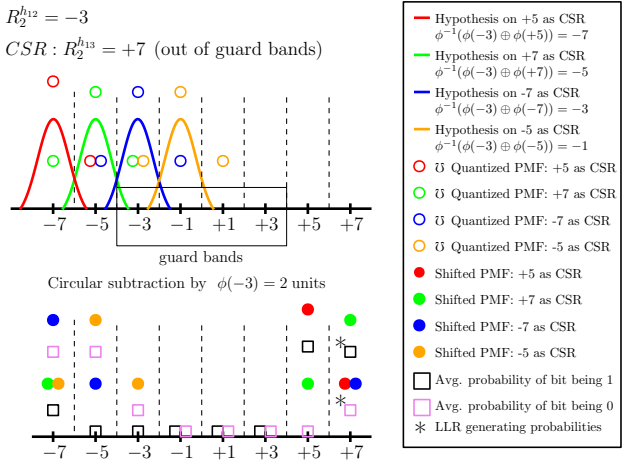}}
\vspace{-0.2cm}
\caption{\label{fig:LLR} Figure depicts an example for LLR generation for $R^{h_{13}}_2$ at node-2. The top figure illustrates the process of generating aposteriori PMFs on the QAM symbols, whereas the bottom figure illustrates the circular shift operation incorporating subtraction over algebraic ring, and also the computation of LLR.
}
\end{figure}

For exposition, we explain the LLR generation scheme at node-2 when the CSR is $\{R^{13}_{1}, R^{13}_{2}, R^{13}_{3}\}$. As a result, using $R^{h_{13}}_{2}$, node-2 generates the LLR on the bit generated at node-1 using $R^{h_{13}}_{1}$. The corresponding quantized version of the channel with node-1, as seen by node-2, is $C^{h_{12}}_{2}$. Henceforth, we use $\hat{\bar{\mathcal{A}}}^+_{I}$ and $\hat{\bar{\mathcal{A}}}^-_{I}$ to represent the positive and negative points in $\bar{\mathcal{A}}_I$, respectively, that are out of guard bands upon quantization using $\mathcal{Q}(\cdot)$.
\begin{theorem}
\label{Theorem_LLR_mixture}
Using the CSR $R^{h_{13}}_{2}$ at node-2, the probability that node-1 quantizes its CSR $R^{h_{13}}_{1}$ to bit $b$, for $b \in \{0, 1\}$, is given in \eqref{equation_theorem:LLR_decoding_1}, where $\mathcal{S} = \hat{\bar{\mathcal{A}}}^-_{I}$ and $\mathcal{S} = \hat{\bar{\mathcal{A}}}^+_{I}$ when $b = 0$ and $b = 1$, respectively. 
\end{theorem}
\begin{IEEEproof}
node-2 intends to build an aposteriori PMF on its samples conditioned on the hypothesis that the CSR $h^{h_{13}}_{1} \in \hat{\bar{\mathcal{A}}}^+_{I}$. This way, node-2 generates the likelihood of CSR $R^{h_{13}}_{1}$ being bit $1$ conditioned on its CSR $R^{h_{13}}_{2}$. Similarly, using all possible cases of $h^{h_{13}}_{1} \in \hat{\bar{\mathcal{A}}}^-_{I}$, it generates the likelihood of CSR $R^{h_{13}}_{1}$ being bit $0$ conditioned on its CSR $R^{h_{13}}_{2}$. Henceforth, throughout this proof, we explain the steps for generating the likelihood of CSR $R^{h_{13}}_{1}$ being bit $1$. Similar steps can be followed to obtain the likelihood of CSR $R^{h_{13}}_{1}$ being bit $0$. Assuming CSR $R^{h_{13}}_{1} = x_{u} \in \hat{\bar{\mathcal{A}}}^+_{I}$, wherein $R^{h_{13}}_{1}$ takes the $u$-th element of $\hat{\bar{\mathcal{A}}}^+_{I}$, for $1 \leq u \leq |\hat{\bar{\mathcal{A}}}^+_{I}|$. node-2 hypothesizes the point broadcast by node-1 as 
$\mu_{x_u}= \phi^{-1}\bigg(\phi(x_u) \oplus \phi(R^{h_{12}}_{2})\bigg)$. Note that it is possible to assume this since at mid-to-high SNR ranges, we have $R^{h_{12}}_{1} = R^{h_{12}}_{2}$. Using the estimate of the channel between node-1 and node-2, the instantaneous SNR at node-2 is $\frac{|h_{12} + e^{(1)}_2|^2}{E_{avg} \sigma^2}$. Therefore, the PDF of the effective noise as seen by node-2 is Gaussian distributed given by $\mathcal{N}\left(\mu_{x_u},\,\frac{E_{avg} \sigma^2}{|h_{12} + e^{(1)}_2|^2}\right)$. Furthermore, since node-2 decodes on the PAM constellation using MAP decoder, the corresponding aposteriori PMF on the PAM points is given by $\mathcal{H}_{R^{h_{13}}_{2}} = \mho\left(\mathcal{N}\left(\mu_{x_{u}}, \frac{E_{avg} \sigma^2}{|h_{12} + e^{(1)}_2|^2}\right), \bar{\mathcal{A}}_{I}\right)$, wherein the PMF $\mathcal{H}_{R^{h_{13}}_{2}}$ is listed on PAM points when enumerated in the increasing order. Finally, since node-2 obtains the CSR $R^{h_{13}}_{2}$ by performing a modulo subtraction on the algebraic ring using $R^{h_{12}}_{2}$, node-$1$ incorporates the corresponding changes in the PMF as $\bar{\mathcal{H}}_{R^{h_{13}}_{2}} = \varnothing^{\phi(R^{h_{12}}_{2})}(\mathcal{H}_{R^{h_{13}}_{2}})$. Thus, node-$2$ generates an aposteriori PMF on the CSR $R^{h_{13}}_{2}$ under the hypothesis that $x_{u}$ was the CSR at node-1. Using the recovered CSR point $R^{h_{13}}_{2}$, node-2 evaluates the probability using the aposteriori PMF as $\varrho^{R^{h_{13}}_{2}}(\bar{\mathcal{H}}_{R^{h_{13}}_{2}})$. Overall, by considering all possible CSR points of $\hat{\bar{\mathcal{A}}}^+_{I}$, the probability that the CSR point at node-1 is quantized to bit 1 is given by $\mbox{Prob}_{1}(R^{h_{13}}_{2})= \sum_{x_{u} \in \hat{\bar{\mathcal{A}}}^+_{I}} \varrho^{R^{h_{13}}_{2}} (\bar{\mathcal{H}}_{R^{h_{13}}_{2}}) \mbox{Prob}(x_{u}),$ where $\mbox{Prob}(x_{u})$ denotes the probability that the CSR $R^{h_{13}}_{1}$ takes the value $x_{u}$. Along the similar lines, the probability that the CSR point at node-1 is quantized to bit 0 is given by $\mbox{Prob}_{0}(R^{h_{13}}_{2})= \sum_{x_{u} \in \hat{\bar{\mathcal{A}}}^-_{I}} \varrho^{R^{h_{13}}_{2}} (\bar{\mathcal{H}}_{R^{h_{13}}_{2}}) \mbox{Prob}(x_{u})$. Finally, the LLR of the bit at node-1 is given by $\mbox{log}\left(\frac{\mbox{Prob}_{0}(R^{h_{13}}_{2})}{\mbox{Prob}_{1}(R^{h_{13}}_{2})}\right)$. Fig. \ref{fig:LLR} depicts an example for LLR generation at node-2 when it observes $R^{h_{13}}_2$ as the decoded CSR.
\end{IEEEproof}

We highlight that no additional communication-overheads are involved in LLR generation. Furthermore, this method is also optimal at mid-to-high SNR and moderate values of $m$.


\section{Simulation Results}

In this section, we present simulation results on the proposed CSR selection as well as the LLR generation strategy. 

\subsection{Opportunistic Selection of CSR}

\begin{figure}[htbp]
\centerline{\includegraphics[scale = 0.38]{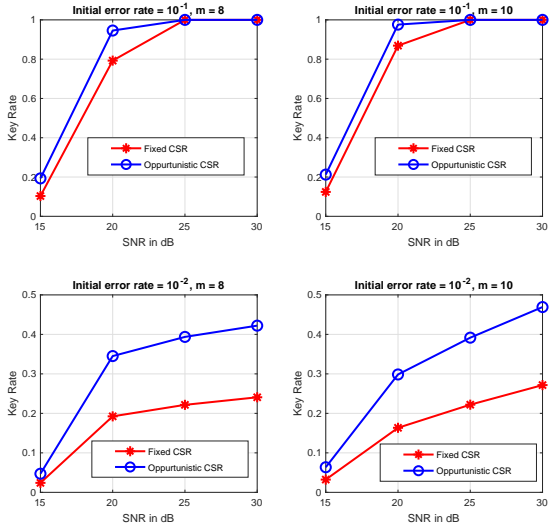}}
\vspace{-0.7cm}
\caption{\label{fig:key_rate_oppt_sel} Key-rate improvement with opportunistic selection of CSR.}
\end{figure}

\begin{figure}[htbp]
\centerline{\includegraphics[scale = 0.38]{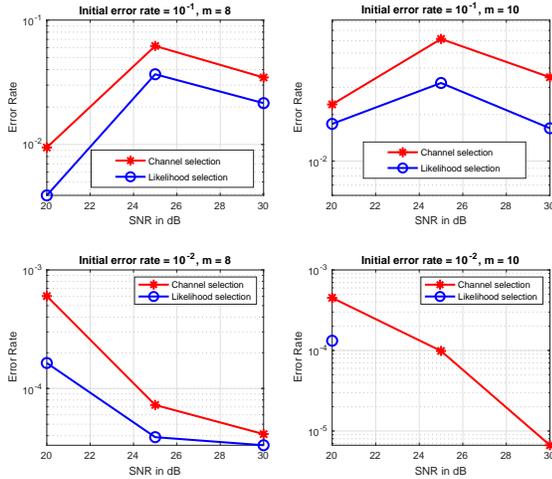}}
\vspace{-0.7cm}
\caption{\label{fig:mismatch_rate_common} Mismatch rate on CSR selected from coherence-blocks when both $\varphi(h_{12})$ and $\varphi(h_{13})$ are in consensus. When the initial error rate is $10^{-2}$ and $m=10$, an error rate of 0 is achieved at $25$ and $30$ dB.}
\end{figure}

To showcase the advantages of the opportunistic CSR selection, we present its key-rate along with that of the A-SQGSK protocol, wherein the CSR is fixed to $\varphi(h_{12})$. In this context, key-rate is defined as the average number of secret bits generated among the three nodes per CSR sample. The plots are presented in Fig. \ref{fig:key_rate_oppt_sel} for the cases when the two-level consensus algorithm delivers secret-keys with an initial error rate of $10^{-1}$ and $10^{-2}$. In this context, initial error rate is defined as the upper bound on the desired mismatch rate among the nodes when choosing the guard bands for the quantizer $\mathcal{Q}(\cdot)$. With an initial error rate of $10^{-1}$, the plots show that the benefits of the opportunistic method is marginal, and this observation is attributed to the fact that the number of samples from $\varphi(h_{12})$ and $\varphi(h_{13})$ that are jointly in consensus is large. However, when the initial error rate is $10^{-2}$, the benefits are significant since the number of additional CSR samples coming out of $\varphi(h_{13})$ is large. We note that the above observations continue to hold good for different values of $m$, which captures the size of the constellation. In this work, we have also proposed a method to choose the CSR sample on those coherence-blocks whenever both $\varphi(h_{12})$ and $\varphi(h_{13})$ are in consensus. To showcase the efficacy of the CSR selection method, in Fig. \ref{fig:mismatch_rate_common}, we plot the error rate offered by our scheme on the CSR samples when both $\varphi(h_{12})$ and $\varphi(h_{13})$ are in consensus.  The plots show that the likelihood based CSR selection outperforms the channel-strength based CSR selection, wherein node-1 chooses the CSR that offers weaker channel-strength since the weaker channel degrades the SNR when recovering the other CSR.

\subsection{LDPC Based Reconciliation for Algebraic-SQGSK}
\begin{figure}[htbp]
\centerline{\includegraphics[scale = 0.38]{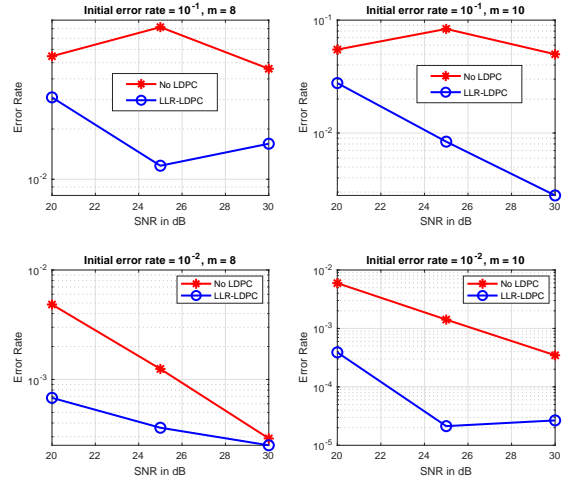}}
\vspace{-0.7cm}
\caption{\label{fig:ldpc_plots} Improvement in the mismatch rate among the keys at the three nodes when LDPC based reconciliation is employed at node-2 and node-3.}
\label{fig:llr_results}
\end{figure}

In the context of opportunistic A-SQGSK protocol, the CSR samples in consensus come from either $\varphi(h_{12})$ or $\varphi(h_{13})$. With respect to the CSR samples from $\varphi(h_{12})$, node-2 and node-3 generate the LLR values on the bits generated at node-1 by following the algorithm at the reciprocal node and the decoding node, respectively. Similarly, for the CSR samples from $\varphi(h_{13})$, node-3 and node-2 generate the LLR values on the bits generated at node-1 by following the algorithm at the reciprocal node and the decoding node, respectively. To execute LDPC based reconciliation, an $(N,K)$ binary LDPC code characterized by the parity check matrix $\mathbf{H}$ of dimension $(N - K) \times N$ is used. With $\mathbf{x} \in \mathbb{F}^{N}_{2}$ denoting an $N$-length binary key generated at node-1, let $\mathbf{s}$ denote its syndrome vector $\mathbf{s} = \mathbf{H}\mathbf{x} \in \mathbb{F}_{2}^{(N-K) \times 1}$, wherein the multiplication operation is over the field $\mathbb{F}_{2}$. Subsequently, the syndrome vector $\mathbf{s}$ is broadcast to node-$2$ and node-$3$, which in turn use it to reduce the mismatch rate by using a message-passing algorithm \cite{R11}. In Fig. \ref{fig:llr_results}, we plot the performance of LDPC reconciliation when an $(N = 12, K = 9)$ LDPC code \cite{R12} is employed. To generate the simulation results, we use the CSR samples out of a two-level consensus algorithm with an initial error rate of $10^{-1}$ and $10^{-2}$. Upon using LDPC reconciliation, we observe that the mismatch rate among the nodes reduces significantly. It is important to note that the benefits of the reconciliation algorithm is attributed to the LLR generation method at the reciprocal node and the decoding node. While we see significant improvements in the mismatch rate in the GSK, we believe that with the use of large block-length LDPC codes, our LLR generation method can ensure zero mismatch rate among the nodes. From the plots, we also remark that the error rate values at $25$ dB and $30$ dB are more than that at $20$ dB when the initial error rate is $10^{-1}$, and this is because the initial error rate is only used as an upper bound.

\section{Conclusion}

In this work, we have presented an opportunistic CSR selection scheme to achieve a higher key-rate than the state-of-the-art A-SQGSK scheme when synthesizing a GSK in a three-node network. Towards guaranteeing non-zero leakage of the CSR to an eavesdropper, we have shown that the proposed CSR selection strategy picks the CSR that minimizes the mismatch rate between the nodes. Finally, to facilitate information reconciliation on the proposed opportunistic CSR selection scheme, we have proposed a novel LLR generation scheme that exploits the underlying noise statistics at the nodes as well as the algebraic ring structure.

\bibliography{reference} 
\bibliographystyle{ieeetr}

\end{document}